**Exploring Strategies for Personalized Radiation Therapy: Part II – Predicting Tumor Drift Patterns with Diffusion Models**

Hao Peng, Steve Jiang and Robert Timmerman

Department of Radiation Oncology, The University of Texas Southwestern Medical Center, Dallas, TX 75390, USA.

**Corresponding Author:**

Hao Peng, PhD, Email: Hao.Peng@UTSouthwestern.edu

**Running title:** Personalized Radiation Therapy

The authors have no conflicts to disclose.

Data sharing statement: The data that support the findings of this study are available from the corresponding authors upon reasonable request.

Author Contributions: Hao Peng: Methodology, Software, Formal Analysis, Investigation, Writing-Original Draft, Review & Editing. Steve Jiang and Robert Timmerman: Methodology, Review & Editing.

**Abstract**

Radiation therapy outcomes are decided by two key parameters—dose and timing—whose best values vary substantially across patients. This variability is especially critical in the treatment of brain cancer, where fractionated or staged stereotactic radiosurgery improves safety compared to single fraction approaches, but complicates the ability to predict treatment response. To address this challenge, we employ Personalized Ultra-fractionated Stereotactic Adaptive Radiotherapy (PULSAR), a strategy that dynamically adjusts treatment based on how each tumor evolves over time. However, the success of PULSAR and other adaptive approaches depends on predictive tools that can guide early treatment decisions and avoid both overtreatment and undertreatment. However, current radiomics and dosiomics models offer limited insight into the evolving spatial and temporal patterns of tumor response. To overcome these limitations, we propose a novel framework using Denoising Diffusion Implicit Models (DDIM), which learns data-driven mappings from pre- to post-treatment imaging. In this study, we developed single-step and iterative denoising strategies and compared their performance. The results show that diffusion models can effectively simulate patient-specific tumor evolution and localize regions associated with treatment response. The proposed strategy provides a promising foundation for modeling heterogeneous treatment response and enabling early, adaptive interventions—paving the way toward more personalized and biologically informed radiotherapy.

# Introduction

Radiation therapy is governed by two fundamental variables—time and dose—both of which should be optimized taking into account substantial inter-patient variability, but remain underutilized in personalizing treatment. This gap is particularly evident in the management of brain tumor, where understanding how these parameters shape treatment outcome is critical. Conventional radiotherapy often employs uniform fractionation schemes (e.g., 2 Gy per fraction), but this one-size-fits-all approach may not reflect individual tumor dynamics. Stereotactic radiosurgery (SRS), particularly in single-fraction form, has demonstrated high rates of local control. However, its efficacy is tempered by an increased risk of neurotoxicity in larger tumors [1–4]. To address this, alternative dosing strategies such as fractionated SRS (FSRS) and staged SRS (SSRS) have been developed, aiming to strike a safer balance between efficacy and toxicity [5-7]. These evolving paradigms underscore a pressing need to move from standard protocols toward data-driven personalization.

In FSRS, high doses (e.g., 6-10 Gy per fraction) are administered over a span of three to five days. One study examined 102 patients treated with FSRS, and showed no significant difference in local control or survival compared to SRS but highlighted a significant benefit in terms of reduced neurotoxicity [7]. In a study involving 214 patients with brain metastases (BM), single-fraction SRS demonstrated comparable response rates between radio-resistant and radio-sensitive tumors. However, FSRS was found to be less effective in treating radio-resistant tumors, highlighting potential limitations of fractionation in this subgroup. SSRS extends treatment over 2-4 weeks to further enhance local control while minimizing adverse effects. Studies have shown varied results, with some reporting good local control and reduced neurological deficits, while others highlight different tumor volume reductions and survival outcomes [8-12].

At our institution, we implemented a special form SSRS: personalized ultra-fractionated stereotactic adaptive radiotherapy (PULSAR) [13-17]. This strategy offers enhanced adaptability based on evolving tumor characteristics and the combination of immunotherapy to explore the synergy between the two. The necessity of personalization in radiation therapy, including PULSAR, is underscored by each patient responding differently to treatment. Some patients experience tumor reduction while others show tumor growth. Balancing tumor control with normal tissue sparing adds another layer of complexity to the choices of dose and timing. To fully take advantage of the PULSAR approach, early and informed decision-making is essential helping to avoid both overtreatment and undertreatment. However, most prior research has focused on outcome prediction using multi-omics data (e.g., radiomics, dosiomics) through classification or regression models [18-26]. While these approaches offer valuable insights, they largely overlook the spatial dynamics of tumor evolution. To be more specific, radiomics can effectively classify tumors based on global imaging features, but it lacks the granularity to predict how the tumor shifts or transforms in space over time in response to therapy. This limitation constrains our ability to track tumor progression at a spatially resolved level—precisely the type of insight needed to guide adaptive treatment decisions.

Building on our Part I study, which used Class Activation Mapping to identify tumor sub-regions associated with treatment outcomes, Part II aims to develop a model that predicts tumor response by capturing the spatial evolution of tumor dynamics. To achieve this, we employ Denoising Diffusion Implicit Models (DDIM) models [27], which learn structured transformations from pre-treatment to post-treatment images by gradually adding noise during training and removing it during inference. Compared to Denoising Diffusion Probabilistic Models (DDPM) [28], DDIM offers one advantage: its non-Markovian forward process enables faster sampling. We hypothesize that this denoising framework can learn a direct, data-driven mapping from a tumor's initial state to its treatment outcome. By anchoring the final target on the post-treatment image during training, the model implicitly refines each denoising step to reflect realistic patterns of tumor evolution. During inference, it can generate plausible outcomes—effectively answering the question of "what could happen", offering a probabilistic view of individualized tumor trajectories.

**Results**

**Raw PULSAR data.** We analyzed preliminary results from a cohort of patients with brain metastases (BMs) treated using the PULSAR approach. A total of thirty-nine patients were included in this study, including 14 males and 25 females with a median age of 61 years (range: 28 to 84 years). This cohort included a set of 69 BM lesions, of which 26 were single metastases and 43 were multiple metastases. The gross tumor volume (GTV) varied between 14.5 mm$^3$ and 37607.6 mm$^3$. Details about patient enrollment and imaging timelines are provided in **Table S1**. As shown in **Figure 1**, the interval between treatment cycles is extended to allow for a second MRI, enabling response assessment. Our earlier publication describes comprehensive analyses using standard radiomics and dosiomics features [29, 30]. Some lesions show a reduction in GTV, while others show continued growth or progression, highlighting the heterogeneity of treatment outcomes. In this study, we focused on the spatial drift between the first MRI and the second MRI. The presence of less drift may indicate radio-resistance, due to variations in anatomical structure, oxygen levels, blood supply, biological traits, and genetic makeup.

**Diffusion model and drift prediction.** We designed an AI framework that combines a U-Net backbone with conditioning to model the denoising trajectory across diffusion steps. The time input is embedded through a fully connected layer and fused with image features at the U-Net bottleneck to condition the network at each diffusion stage. The image is processed through standard U-Net encoding and decoding layers, with skip connections to help preserve spatial detail. To improve learning, intermediate layers are shared between the noises and drift prediction branches. The model takes a noisy pre-treatment image and a scalar time step as input and outputs either the predicted post-treatment image (Method 1) or the predicted noise (Method 2) along with a drift map, as shown in **Figure 2**. Method 1 functions like a standard autoencoder under different noise levels, guiding the output towards a fixed clinical endpoint in just a *single* step. Method 2 follows standard DDIM, where the model learns to progressively remove noise step by step. In both methods, the drift map serves as an interpretable output, highlighting regions of meaningful change due to treatment. In essence, by training with real post-treatment data, the model learns to reverse noise not just into any plausible image, but into outcome-consistent images.

**MNIST zero digit as a toy example.** We began our experiments by testing the model on the MNIST dataset, focusing specifically on the digit "0". To simulate lesion growth or shrinkage, the zero digits are scaled isotropically—maintaining their aspect ratio by equally adjusting both height and width. **Figure 3** illustrates how the diffusion model processes images over multiple steps, for three scenarios: Method 1, Method 2, and one fully decoupled model (i.e., predict the post-treatment image and the drift field separately). All three methods show satisfactory performance in reconstructing zero digits. The third one performs comparably to Methods 1 and 2. On one hand, incorporating noise into training acts as a form of regularization, encouraging the model to learn smoother and more meaningful feature representations. On the other hand, denosing helps the model to recover critical features in terms of shape, boundaries, and texture. Nonetheless, applying this approach to real tumors will be much more challenging due to their complex structures and heterogeneous drift behaviors.

**Data augmentation, loss function and model training.** Data augmentation was performed by rotating the lesion at four different angles to increase sample diversity (**Fig. 4A**). Noise was added at 20% of the largest signal intensity to keep a reasonable signal-to-noise ratio (SNR) (**Fig. 4B**). Variations in tumor size and the presence of many zero-valued pixels posed challenges for the U-Net. To address this, a bounding box was used to first mask the GTV and drift regions and then rescaled to a standard size for all lesions. During training, loss curves for both image reconstruction and drift prediction were monitored. The model was trained using a combination of mean squared error (MSE) loss and structural similarity index (SSIM) loss, weighted at different ratios. While MSE focuses on pixel-level accuracy, SSIM emphasizes structural features such as edges and textures, which help preserve perceptual image quality. Training was conducted over approximately 5,000 iterations, with significant improvements observed within the first 1,600 iterations and further refinement occurring during the final 800 iterations (**Fig. 4C**).

**Method 1. Figure 5** shows the performance of a single lesion with varying weight ratios for reconstruction and drift: 1:1, 1:1.5, 1:3.0, along with the pre-treatment image, post-treatment image, and ground truth drift. The reconstruction loss combines 90% MSE and 10% SSIM to balance pixel accuracy with structural integrity. For drift prediction, only MSE is used. A noise level of 0.2 was applied during training. **Figure 5C** achieves relatively better predictions than its counterparts, preserving both central features and edges. It is also observed that at higher noise levels (early steps), the model recovers coarse structural information. Midway through the denoising process, the model reaches a "sweet spot"—most of the noise has been removed, but fine details are still preserved. On the right end when little noise stays, the model makes only subtle adjustments, and over-smooth the image. This over-smoothing is likely influenced by the linear noise schedule: while the denoising progresses at a steady pace, the complexity of features to be recovered does not (e.g., different rate of change at each step). Consequently, the model unnecessarily spends too many steps refining an already clean image and blurs fine details. Optimal performance is achieved when the SNR is balanced—enough noise to challenge the model, but not so much that the signal is lost. Increasing the noise level (e.g., to 0.5) in the forward diffusion process degrades performance, and the model does not capture finer drift structures (**Figure S1**).

**Method 2**. The training converges after approximately 1600 iterations (**Figure S2**). Like Method 1, the model in Method 2 struggles to function under poor SNR (50% noise level) (**Figure S3**). At such high noise levels, the reverse diffusion process becomes unstable and cannot reliably distinguish between random noise and meaningful image structure. **Figure 6** illustrates the performance of Method 2 on five lesions, where the network uses separate layers for decoupling noise and drift, and the loss function incorporates both drift-based SSIM and MSE. The model can progressively denoise and reconstruct the drift map. Unlike Method 1, Method 2 shows a more stable drift map, providing intermediate "snapshots" of progression. During each reverse step, Method 2 only applies small corrections, allowing for a more delicate restoration and preserving fine details. In contrast, one-step denoising in Method 1 must reconstruct the entire image from a severely distorted state in a single leap.

Based on visual inspection, the performance shows only slight improvement compared to the two counterparts: shared layers (**Figure S4**) and MSE-only loss (**Figure S5**), evaluated on the same five lesions. Given the complex drift patterns, incorporating SSIM into the loss function noticeably aids in learning more consistent alignments. In multi-loss training setups, if one loss term (e.g., reconstruction loss) dominates while another (e.g., drift loss) is relatively small, the weaker loss may not effectively guide the model. In our study, such imbalance was found to destabilize the reverse process when not properly optimized. Moreover, competing objectives—such as noise removal and drift modeling—also interfere with each other, especially when not using two fully decoupled decoder layers. Inaccuracies in noise estimation can propagate into the drift component, further compounding reconstruction errors.

**Figures 7** and **8** illustrate the performance of Method 2 on lesions with poor or good response, using a 20% reduction in tumor volume as the threshold for classification [31-33]. In the drift maps, red regions show higher drift, corresponding to areas of rapid tumor shrinkage, while blue regions show lower drift, suggesting poor or minimal response. The model predicts drift patterns that align well with the true drift map, capturing key spatial changes associated with treatment response. Interestingly, there was no clear difference in drift patterns between responding and non-responding lesions at the group level, suggesting that treatment response is not solely decided by core or peripheral tumor regions. Notably, only a noisy version of the pre-treatment image is needed at the beginning of the inference phase, as the denoising steps have been pre-trained to learn patterns consistent with actual post-treatment changes. Due to the limited sample size in this study, no quantitative evaluation has been conducted yet.

**Discussion**

Personalized radiation therapy goes beyond classification. Given a tumor image right before treatment, over time, some parts grow, others shrink, borders get irregular, etc. Through diffusion processes, the model is trained to progressively learn the mapping between the pre-treatment and post-treatment images, a structured transformation that can capture intricate and nonlinear relationships. By incorporating guidance from both post-treatment image and drift, the denoising is not performed blindly—it allows the model to generate outputs that reflect possible treatment responses, such as tumor shrinkage.

When noise is added to an image, it is distributed randomly, while underlying anatomical structures (e.g., tumor boundaries or tissue interfaces) follow consistent, spatially coherent patterns. During training, the model learns to distinguish between these random noises and meaningful spatial features. Meanwhile, each step of denoising teaches the model what "looks like" a post-treatment image, such as shrinkage in the center, necrotic edges, or other irregular patterns. Adding noise during training encourages the model to learn a hierarchy of features—from low-level details like edges and textures to high-level patterns such as shape—thereby improving its ability to generalize to new data. In the early steps (e.g., higher noise levels), the model focuses on preserving the overall spatial layout and major structures. In later steps (e.g., lower noise levels), with most noise removed, the model fine-tunes fine-grained details, enhancing local features like edges, textures, and small structures. These hierarchical features are implicitly linked to different patterns of response. Another advantage of adding noise is its capability to simulate variability and uncertainty, given sufficient samples available. With different random seeds applied at the beginning of the interference phase, DDIM allows us to model this uncertainty and predict multiple plausible drift patterns, not just one. It is noteworthy that our notion of "drift" is not strictly a physical or temporally continuous process. Since our framework only used two time points (pre- and post-treatment images), we did not explicitly model tumor evolution over time. Our goal is not to reconstruct this entire temporal trajectory. Rather, we extract features most predictive of treatment response.

This study is a first exploratory effort, and a key limitation lies in the relatively small dataset size used for model training. We trained our diffusion model on 828 paired pre- and post-treatment images, derived from 69 lesions (3 representative slices per lesion and 4 augmentations per slice). To partially address the limited sample size, we used 2D slices instead of full 3D volumes and applied basic augmentation strategies such as rotation. However, this dataset is still small for training a diffusion model from scratch, which typically benefits from larger-scale data. The adequacy of the sample size is closely tied to the diversity of tumor response: greater heterogeneity demands more training data to capture the full range of dynamics, while more uniform responses—such as central shrinkage—may allow the model to learn meaningful patterns with fewer examples. Nonetheless, future work will focus on validating and expanding this approach with larger PULSAR datasets and applying it to other tumor types beyond brain metastases. For instance, glioblastoma multiforme presents distinct drift patterns compared to brain metastases, owing to its infiltrative growth and less well-defined tumor margins.

Without the noise-driven framework, one would typically rely on direct image-to-image translation models (e.g., GANs, VAEs) [34, 35], where the pre-treatment image is mapped directly to the post-treatment image. While effective for simpler or more linear changes, GAN and VAE models may struggle to capture complex, nonlinear transformations (e.g., spatial drift patterns). The diffusion model, on the other hand, can learn finer, more nuanced image transitions. Signal and noise are inherently related. By learning to remove noise, diffusion models uncover the underlying structure in MRI images—much like how signal processing filters noise to recover the original signal. Both method 1 and method 2 differ from GANs and VAEs in that they do not use an explicit low-dimensional latent space. Instead, they add noise directly to data and learn to reverse this process to recover clean images. While this noise injection resembles a latent representation, it works in pixel space rather than a compressed form. Doing so also leads to more stable training, avoiding issues like mode collapse commonly seen in GANs.

A comparison of the two methods highlights their differences in complexity. Method 1 is relatively straightforward: the model denoises toward a fixed reference image in pixel space. This task is local and deterministic, requiring no understanding of the broader data distribution—just a direct path from noisy input to a known clean output. In contrast, Method 2 is a generative task that works globally and is manifold-aware (i.e., complete distribution of pre-treatment MRI dataset). As shown in **Fig. 5**, Method 1 predicts tumor drift in a few denoising steps, while Method 2 requires a longer reverse trajectory through the diffusion process (**Fig. 6**). One plausible explanation lies in the complexity of the data. Unlike simpler datasets such as MNIST (e.g., digit zero), MRI images contain rich textures and highly variable tumor morphologies. This makes it more difficult to learn the complex, nonlinear mapping between pre-treatment input and post-treatment outcome—especially when working with a limited dataset. In Method 2, the forward diffusion process moves the image off the data manifold, and the reverse process must reconstruct a coherent structure from noise. With only 828 samples, the manifold is sparsely sampled, making it challenging for the model to reliably learn a meaningful and stable reverse trajectory. Despite these limitations, Method 1 currently offers more robust performance. In our view, Method 2 is preferred for its generative capabilities—offering the potential to progressively simulate diverse tumor trajectories. This makes it more aligned with the goal of personalized, probabilistic modeling of tumor evolution. Future work will prioritize improving Method 2, once more data becomes available.

While radiomics and dosiomics can be used for treatment outcome classification, our study focuses on predicting the spatial drift after treatment. We speculate that the presence of less drift may indicate radio-resistance, due to variations in structural heterogeneity. This hypothesis could be evaluated through further investigation to find whether these regions align with underlying tumor biology or treatment-induced pathological patterns. For example, these regions could be compared with pathological or molecular patterns, such as those identified through single-cell resolution CODEX (CO-Detection by indexing) [36-38] or pixel-level spectroscopic MRI (MRS) [39]. For instance, CODEX can provide spatial information about the molecular and cellular makeup of the tumor, such as the expression of specific genes, proteins, and immune markers. This information can help identify tumor heterogeneity and the presence of resistant subpopulations within different regions of the tumor. MRS can help measure metabolic activity (e.g., lactate, choline, lipids) and oxygenation levels within the tumor. These factors are crucial in understanding tumor heterogeneity, as regions with low oxygen levels (hypoxia) or high metabolic activity are often more resistant to radiation therapy. Although MRI and pathology differ vastly in spatial resolution—approximately 0.5 mm for MRI versus 20 microns for pathology—methods correlating them from either broadly corresponding or perfectly co-localized regions have emerged as effective tools for virtual biopsies. For example, Brancato et al. analyzed MRI-derived radiomic features, such as Apparent Diffusion Coefficient (ADC) and T1 contrast, alongside pathomic characteristics in glioblastoma patients [40]. Bobholz et al. employed an approach by co-registering postmortem multi-sequence MRIs directly with histological images in brain cancer patients [41]. Their model predicted histology-derived cellularity from localized MRI intensities, successfully mapping hypercellular regions confirmed by immunohistochemical markers of proliferation and microvascular density. These findings highlight the potential to link imaging features—particularly spatial drift—with specific histological sub-regions, thereby enhancing the biological interpretability of treatment response.

We envision that the framework might find its potential use in two clinical scenarios, towards personalized radiation therapy. First, the proposed framework can be used for simulating treatment outcomes for PULSAR and other fSRS treatments. Being able to simulate how an individual patient might respond to treatment enables dynamic adjustment of the plan as early as possible. Method 2 produces a distribution of possible future states, rather than a single deterministic prediction. Treatment effects are inherently uncertain due to biological variability. Generating several post-treatment images from the same pre-treatment images enables averaging to produce a more robust prediction. If all generated outcomes (e.g., drift map) are similar regardless of noise, the model likely has high confidence in its prediction. However, in case there is substantial variability, it may indicate that the treatment effect is unpredictable for that patient. Second, the proposed framework holds promise for distinguishing between different types of treatment responses. Although not explored in the current study due to the limited sample size, one powerful extension is conditional generation based on response class labels (**Figs. 7** and **8**). By conditioning the diffusion model on labels such as "good response" or "poor response," the model can generate class-specific post-treatment outcomes, even in the presence of stochastic noise. For example, conditioning on a "good response" label would allow the model to produce multiple plausible variations of favorable treatment outcomes, helping to illustrate the spectrum of successful tumor shrinkage. Conversely, conditioning on a "poor response" label would yield outcomes that reflect minimal regression or progression, such as persistent or enlarging tumor regions. This capability may offer some useful insights for clinicians seeking to understand the likely trajectories of different response types.

**Conclusion**

Radiation therapy should not be a one-size-fits-all approach. Our study highlights the potential of diffusion models to predict spatial tumor drift, capturing patient-specific treatment dynamics. By developing and comparing single-step and iterative denoising strategies, we showed that diffusion models can effectively simulate individualized tumor evolution and identify spatial drifts after treatment. This approach lays a promising foundation for modeling heterogeneous responses and helping early, adaptive interventions—advancing the goal of more personalized, biologically informed radiotherapy.

**Method**

***Problem Formulation.*** We treat tumor progression as a diffusion process. Given a tumor image right before treatment, can we use a learned model to predict what the tumor will look like after treatment? Over time, some parts grow, others shrink, borders get irregular, etc. The drift field is a map that tells each pixel where to move, how to change intensity, or what new structure to form, representing the direction and strength of progression.

$$Post\ Image\ =\ Pre\ Image + drift \qquad (1)$$

In the forward diffusion process (only use pre-treatment image, represented by $X_0$), we progressively add noise to the image following a pre-defined linear noise schedule:

$$X_t = \sqrt{\bar{\alpha}_T}X_0 + \sqrt{1-\bar{\alpha}_T}\ \varepsilon_0 \qquad (2)$$

$X_t$ is the noisy image at time t. During the reverse (denoising) process, the model is trained to reconstruct noise term $\varepsilon_\theta(X_t, t)$ and drift term $\mu_\theta(X_t, t)$.

$$\varepsilon_\theta(X_t, t), \mu_\theta(X_t, t) = UNet(X_t, t)$$

$$Loss_{recon} = L\|\varepsilon - \varepsilon_\theta(X_t, t)\|^2 \qquad Loss_{drift} = E_t\|drift - \mu_\theta(X_t, t)\|^2 \qquad (3)$$

Since only two images (pre- and post- treatment) are available in our task, the drift is not modeled as time-dependent in our framework. From a biological perspective, drift is likely to vary over time—for instance, a gradually shrinking tumor would show distinct drift patterns at different stages. For the time being, our implementation does not explicitly model tumor evolution as a continuous *temporal* process. Instead, the focus is to use the model to extract the most informative features associated with treatment response. By anchoring the output on the target (the post-treatment image), those features capture underlying biological pathways or contextual factors driving the response to radiation.

During the reversion process, the two methods adopt distinct approaches. Method 1 functions as a standard autoencoder, predicting drift across varying noise levels. Method 2, however, uses the DDIM framework to model the diffusion trajectory as shown below. At each step, the model predicts $\varepsilon_\theta(X_t, t)$. Once $\varepsilon_\theta(X_t, t)$ is known, we calculate $\hat{X}_0$.

$$\hat{X}_0 = \frac{X_t - \sqrt{1-\bar{\alpha}_T}\,\varepsilon_\theta(X_t, t)}{\sqrt{\bar{\alpha}_T}} \qquad (4)$$

For the reverse sampling, the key feature of DDIM is that instead of sampling the next noisy image stochastically like in DDPM. The next sample $X_{t-1}$ in DDIM is:

$$X_{t-1} = \sqrt{\bar{\alpha}_{T-1}}\hat{X}_0 + \sqrt{1-\bar{\alpha}_{T-1}}\,\hat{\varepsilon}_t(X_t, t) \qquad (5)$$

Unlike DDPM, where randomness is reintroduced at each step, DDIM smooths out the denoising process deterministically by controlling the ratio between the two terms, not requiring strict Markov chain assumptions. The first term shifts the next sample closer to the clean image $X_0$, while the second term retains a bit of the noise based on the noise schedule.

***PULSAR data collection.*** PULSAR patients with BMs were treated utilizing Gamma Knife Icon™ (Elekta AB, Stockholm, Sweden). Patients first undergo a pretreatment MRI scan followed by the initial treatment course, which consists of three fractions or pulses (5 to 6 Gy per fraction/pulse) with a two-day interval between fractions. Subsequently, after a span of three weeks, the second treatment cycle is administered based on the second MRI scan, allowing for adjustments to changes in tumor volume. Our retrospective study focused on 39 patients who underwent PULSAR treatment at UTSW. This cohort included 69 lesions treated between November 1, 2021, and May 1, 2023, encompassing patients with both single and multiple metastases. We collected MRI images and radiotherapy contour structure files (RTstructure). The MRI images were acquired using axial (AX) sequences with T1-weighted enhancement. Tumor volumes in follow-up MRI images (without enhancement) were assessed three months after PULSAR treatment. Two board-certified radiation oncologists conducted a thorough comparison of MRI images and GTV contouring for each lesion. Previous studies have demonstrated a strong correlation between a volume reduction of 20% or more and improvement in neurological signs and symptoms [26-28]. We applied the same criterion and framed it as a classification problem. Tumors with follow-up volumes at or above 80% of their initial volume were categorized as "poor response" (referred to as Group A), while those with volumes reduced below this threshold were categorized as "good response" (referred to as Group B). Before evaluating the model on the real ULSAR dataset, we use the zero digit from the MNIST dataset as a toy example to assess feasibility.

***Diffusion model architecture and drift prediction.*** To model the diffusion trajectory, we designed a convolutional neural network that integrates a U-Net backbone with explicit conditioning and an auxiliary drift field prediction head. The model takes a noisy image (pre-treatment) as input and returns two outputs—post-treatment (Method 1) or noise (Method 2), plus a learned drift field. The architecture begins with two input branches: a standard 2D image input and a one-dimensional scalar input representing the time step t. The scalar time input is first embedded into a higher-dimensional representation using a fully connected layer with 64 hidden units and ReLU activation. This encoding is later broadcast and concatenated with the spatial feature maps at the bottleneck of the U-Net to condition the model on the current diffusion step. The image input is passed through a U-Net encoder consisting of convolutional and max-pooling layers to progressively extract hierarchical features. The decoder reconstructs the image using transposed convolutions and skip connections from corresponding encoder layers, enabling the recovery of fine-grained details. This reconstruction path culminates in a final convolutional layer that outputs either the predicted post-treatment image (Method 1) or predicted noise (Method 2). For the second output head (drift field predictor), we tested using either shared convolutional layers or fully decoupled layers to generate a per-pixel vector field representing the estimated "drift" direction.

*Method 1 vs Method 2.* The goal of method 1 is to predict the post-treatment image, denoising with different noise levels towards a fixed target. It acts as a denoising autoencoder-style setup, focusing on learning to map noisy inputs to a known output ("one-to-one mapping"). It is essentially a denoising task, not generation. By contrast, the method 2 based on DDIM starts from a noisy latent and perform step-by-step denoising to generate a clean image. From different noise seeds, DDIM can generate many possible outputs. In Method 1, the model is effectively learning to reconstruct the clean image directly. This gives clear, interpretable supervision as the target is a clean image, making the learning process more straightforward and efficient. In Method 2, predicting the noise is more challenging because of its stochastic nature (i.e., more variability compared to image), as well as balancing noise and drift at the same time.

*AI model development and training.* Data augmentation was introduced by rotating the raw image (69 lesions, each with three slices) at different angles (0°, 90°, 180°, 270°), yielding a total of 828 samples (69×3×4=828). To simulate the forward diffusion process, we applied a linear noise schedule defined as linspace (0.05, 0.95, T), where T=21 denotes the total number of discrete time steps. At each step t, the clean image $X_0$ is perturbed with additive Gaussian noise, resulting in a progressively noisier version of the image. The loss function includes two components. The first part is based on a reconstruction of the post-treatment image (Method 1) or noise (Method 2), with the reconstruction loss combining structural similarity (SSIM) and mean squared error (MSE). The second part is a constant drift vector field representing the difference between pre- and post-treatment MRI images, based on MSE loss. This formulation encourages the network to generate outputs that are structurally faithful to the ground truth, while capturing the underlying directionality imposed by the diffusion process. The AdamW algorithm with default settings was used for optimization (batch size=64, learning rate = 0.0001, weight decay = 0.0001, beta1 = 0.9, beta2 = 0.999, epsilon = $1×10^{-8}$).

42.

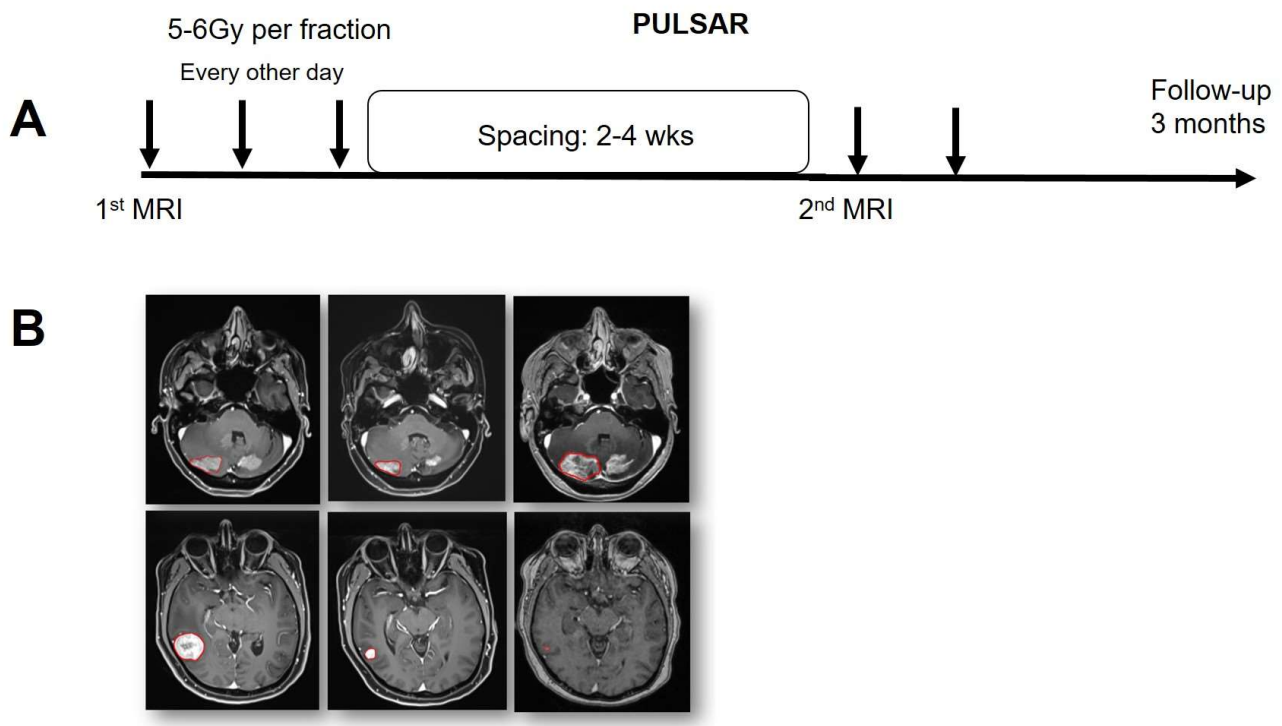

**Figure. 1. A.** Workflow of PULSAR, which extends the interval between two treatment cycles and includes a second MRI. **B.** Lesion volumes trajectories in PULSAR treatment at three time points (first MRI, second MRI, and follow-up). Two examples illustrate increased GTV (top) and decreased GTV (bottom) after treatment.

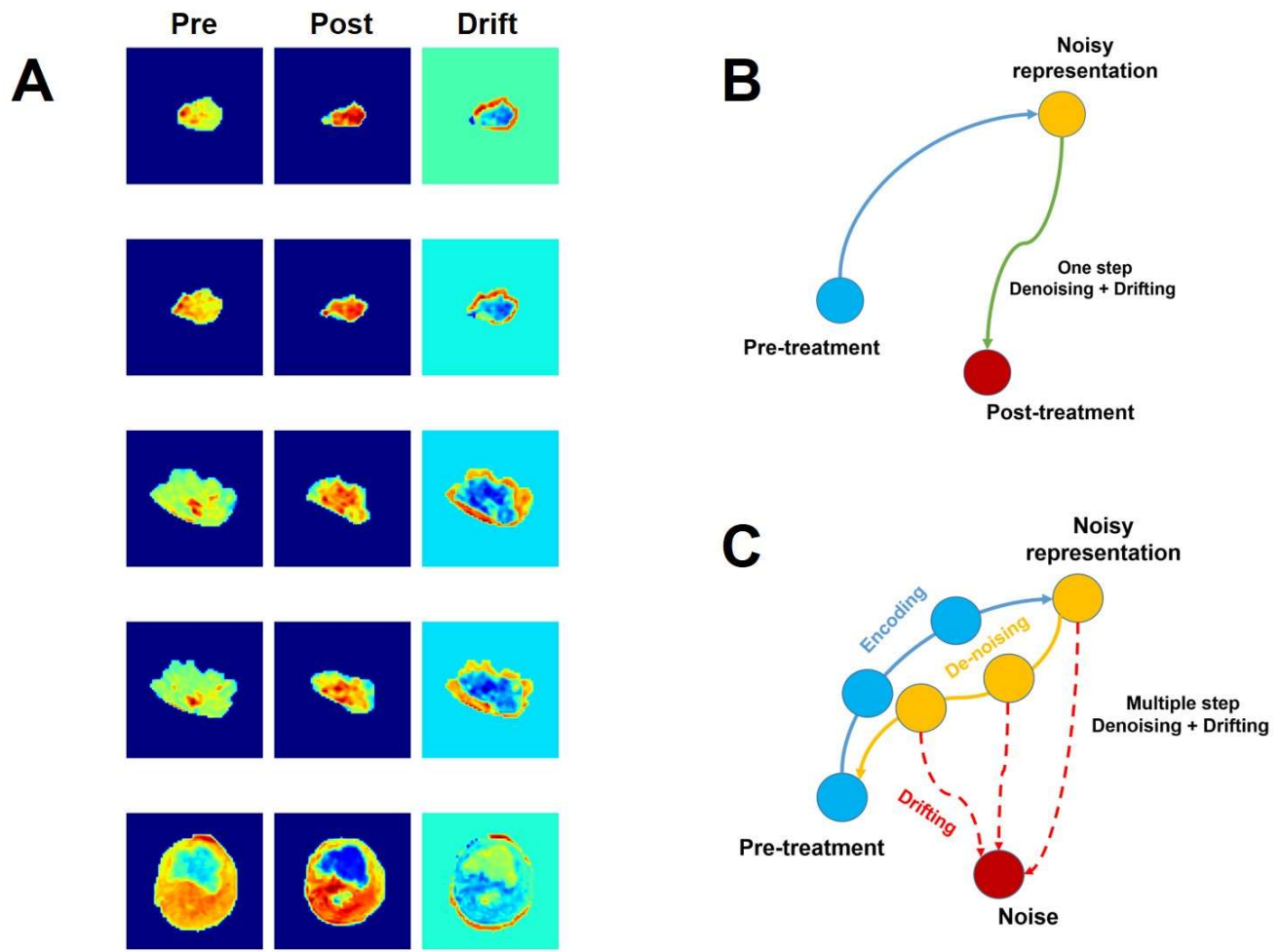

**Figure 2. A.** Six examples showing tumor progression from pre-treatment to post-treatment with the corresponding drift. **B.** Method 1 uses a denoising autoencoder-style approach: it takes a noisy pre-treatment image as input and directly predicts the corresponding post-treatment image along with a learned drift field in a single step. **C.** Method 2 employs a DDIM-based generative model: it begins with a noisy latent input and iteratively denoises it to generate a post-treatment image, also learning the drift field in multiple steps. Unlike Method 1, this approach predicts the noise itself, which introduces stochasticity, allowing the generation of multiple possible outcomes.

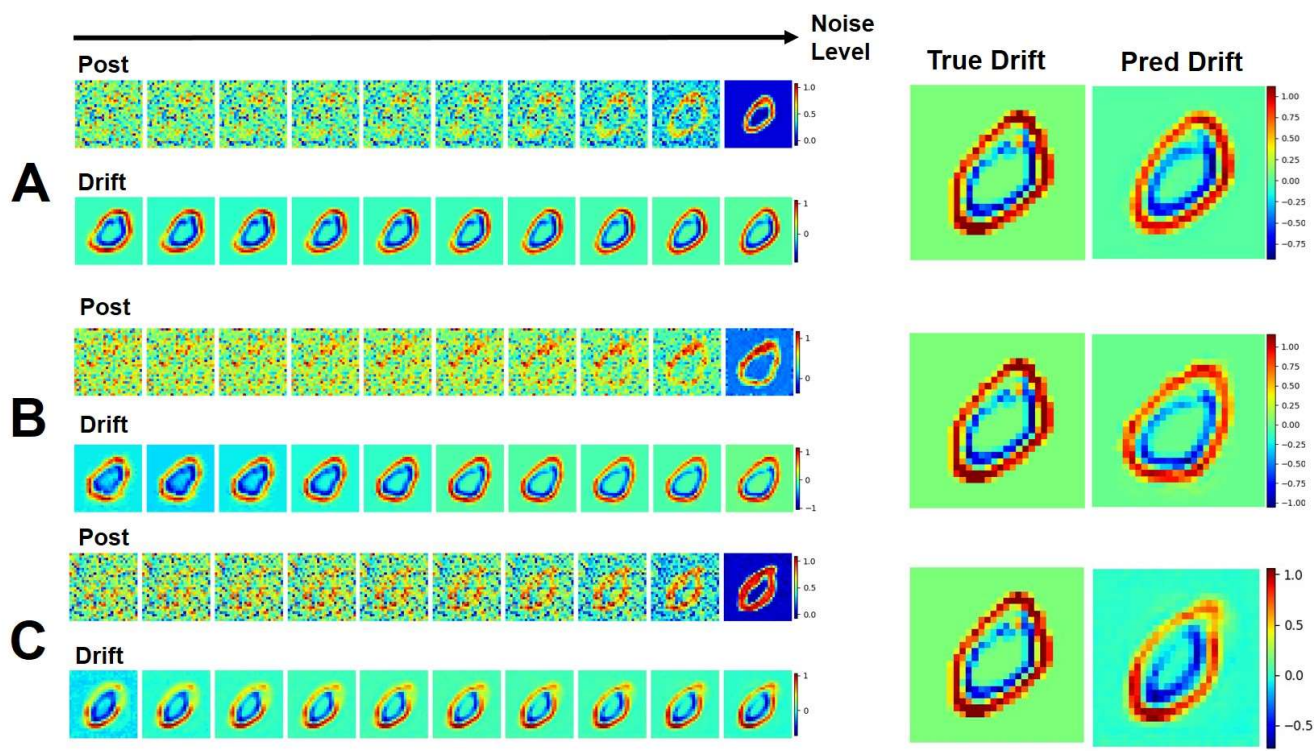

**Figure 3.** Results of the MNIST zero-digit reconstruction task illustrating the diffusion process. **A.** Method 1. **B.** Method 2. **C.** Two fully decoupled networks trained to predict the post-treatment image and the drift field separately. All three methods show satisfactory performance in reconstructing or generating zero digits.

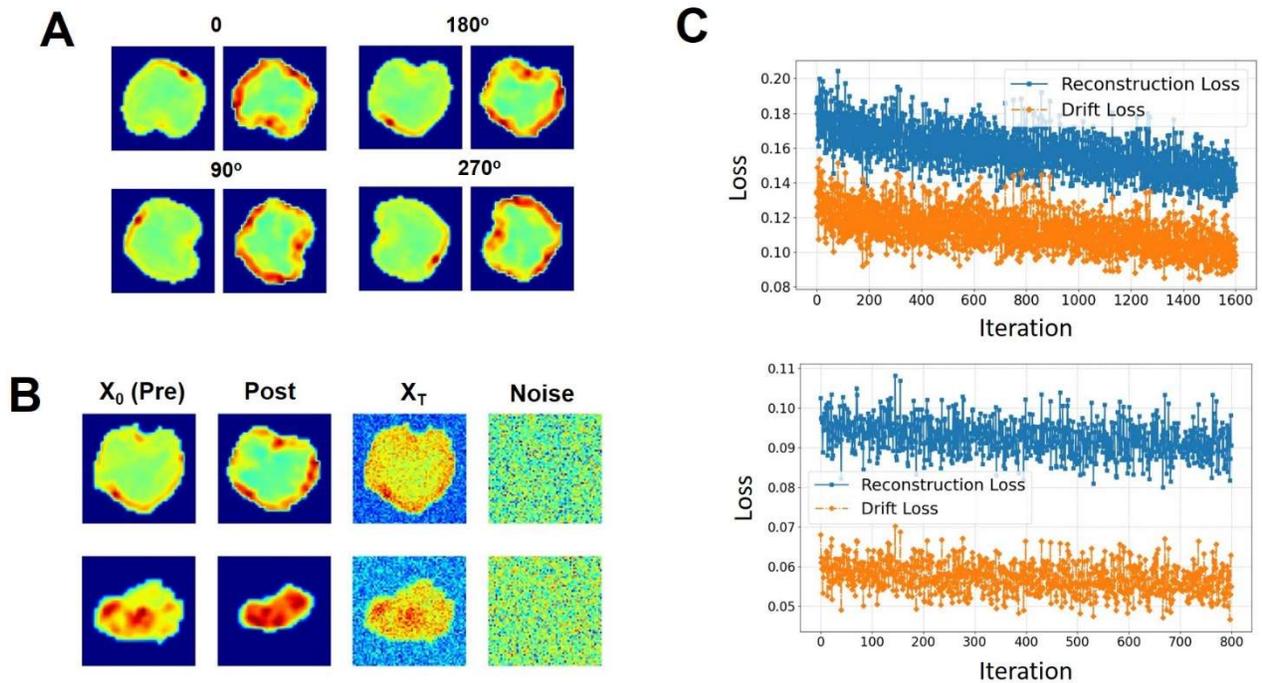

**Figure 4. A.** Data augmentation by rotating the lesion at four different angles to increase sample diversity. **B.** Illustration of the diffusion process and noise addition for two selected examples. $X_0$ represents the pre-treatment image, $X_{post}$ is the post-treatment image, and $X_T$ is the noisy version of X0 after noise has been added. **C.** Loss curves continuously monitored for both reconstruction and drift during training (Method 1). Top: Over the first 1600 iterations. Bottom: Over the final 800 epochs. Training concludes at approximately 5000 iterations.

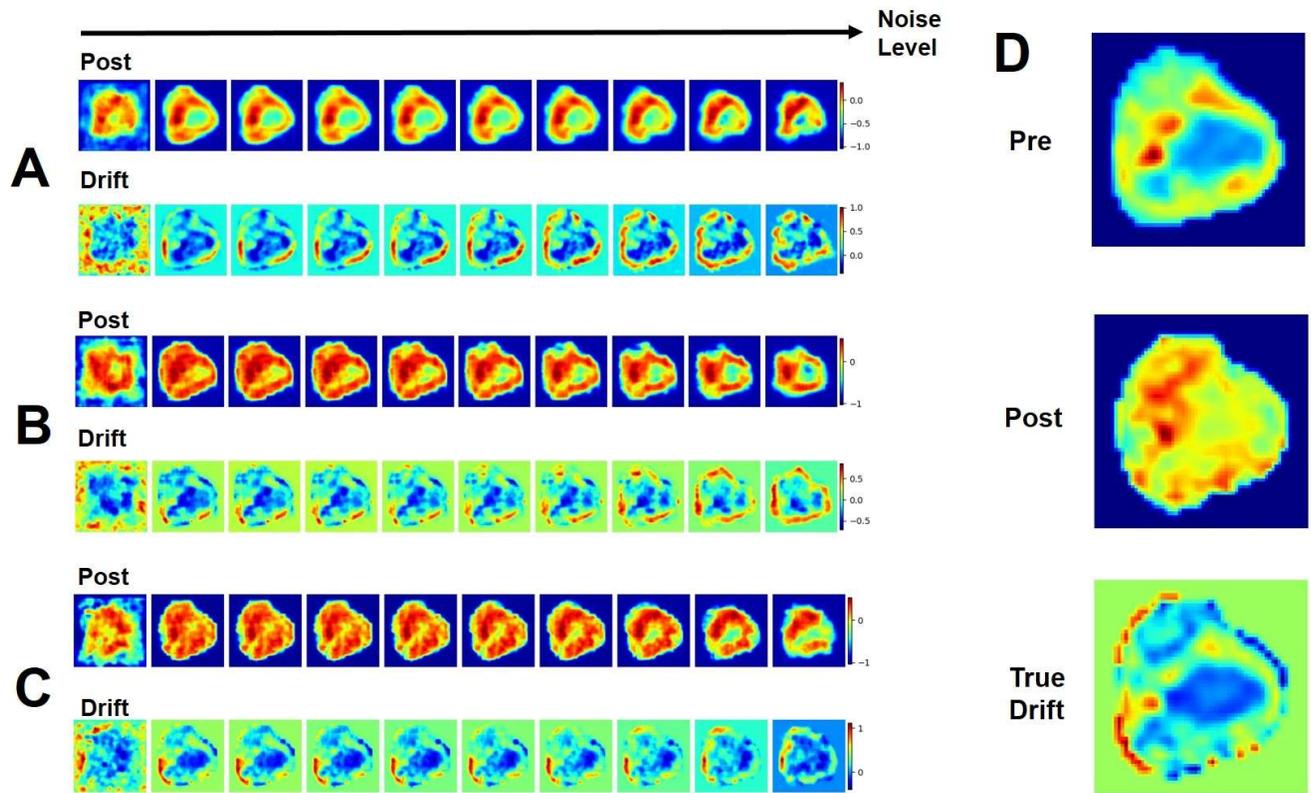

**Figure 5.** Performance of Method 1 on a single lesion with varying weights for reconstruction and drift losses: **A.** Loss weight ratio 1:1. **B.** Loss weight ratio 1:1.5. **C.** Loss weight ratio 1:3.0. **D.** The pre-treatment image, post-treatment image, and the ground truth drift.

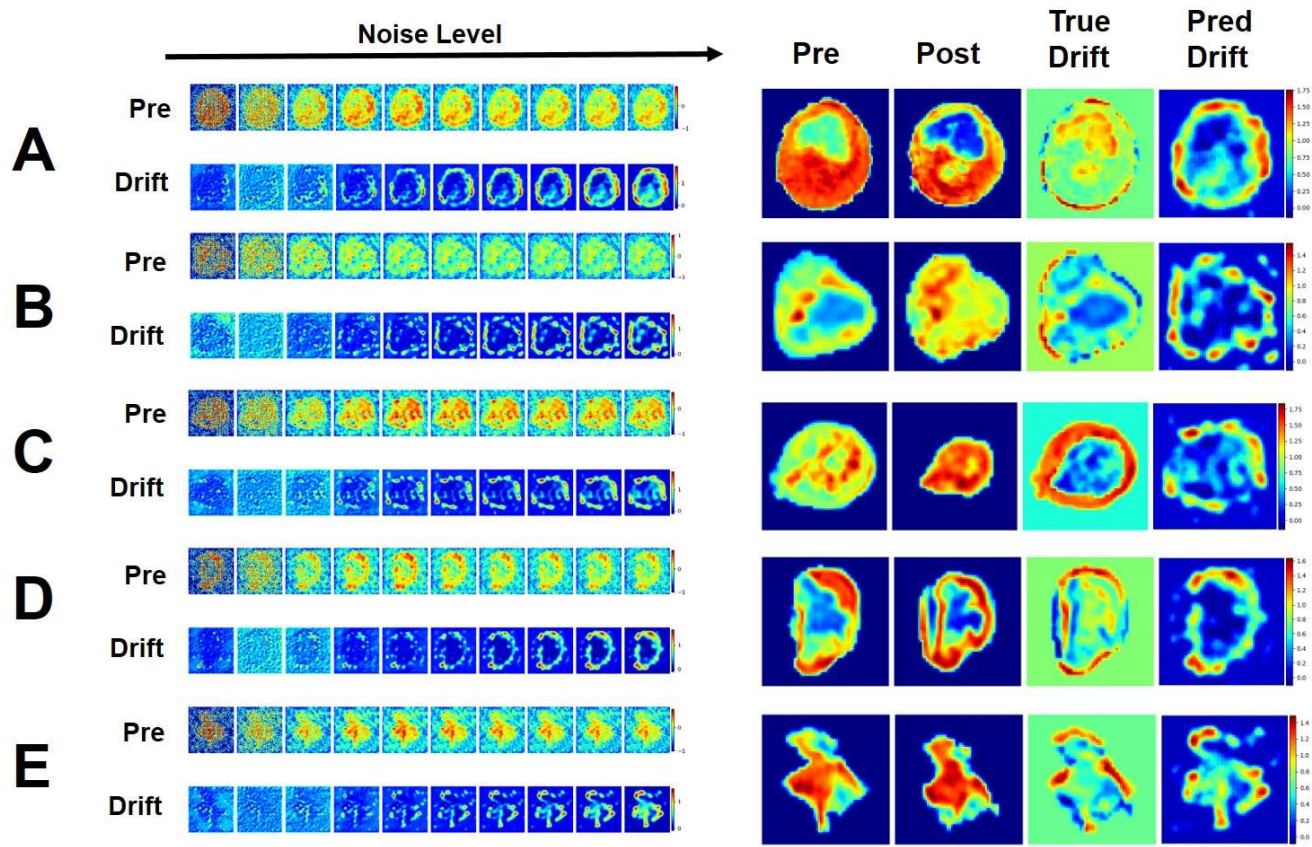

**Figure 6.** Performance of Method 2 on five lesions (A to E), using two decoupled network layers to predict the post-treatment image and the drift field separately, with the drift loss (including SSIM and MSE) as weak guidance.

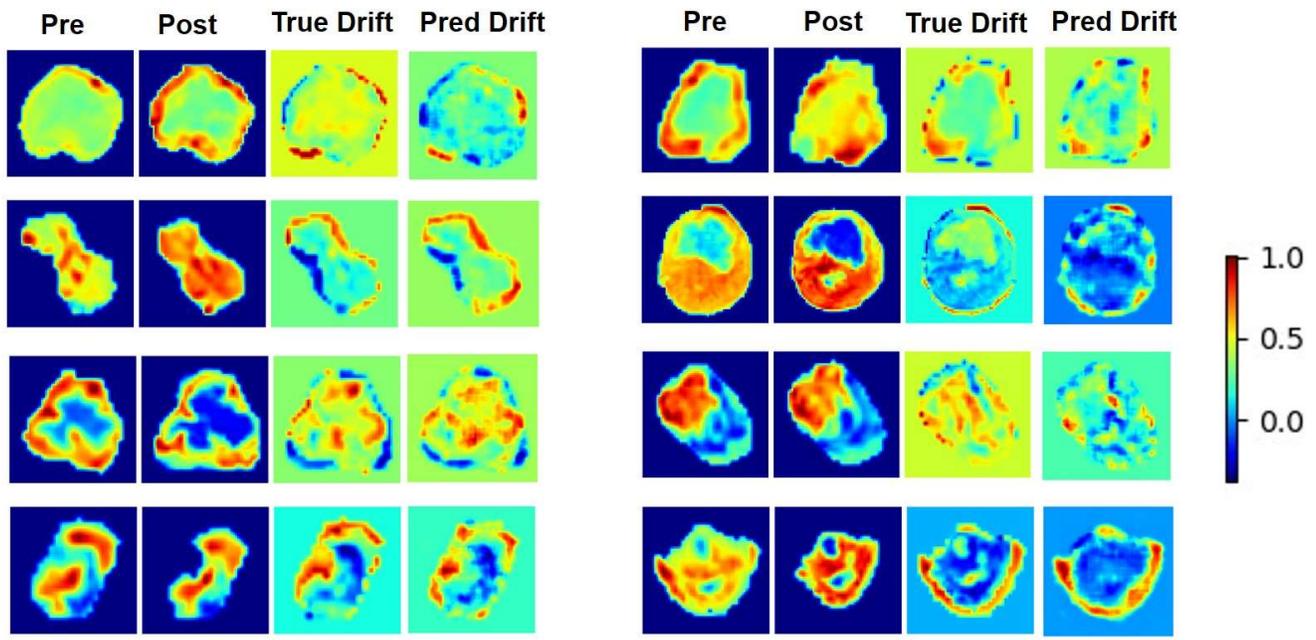

**Figure 7.** Performance of Method 2 for lesions showing poor response (tumor volume reduction less than 20%).

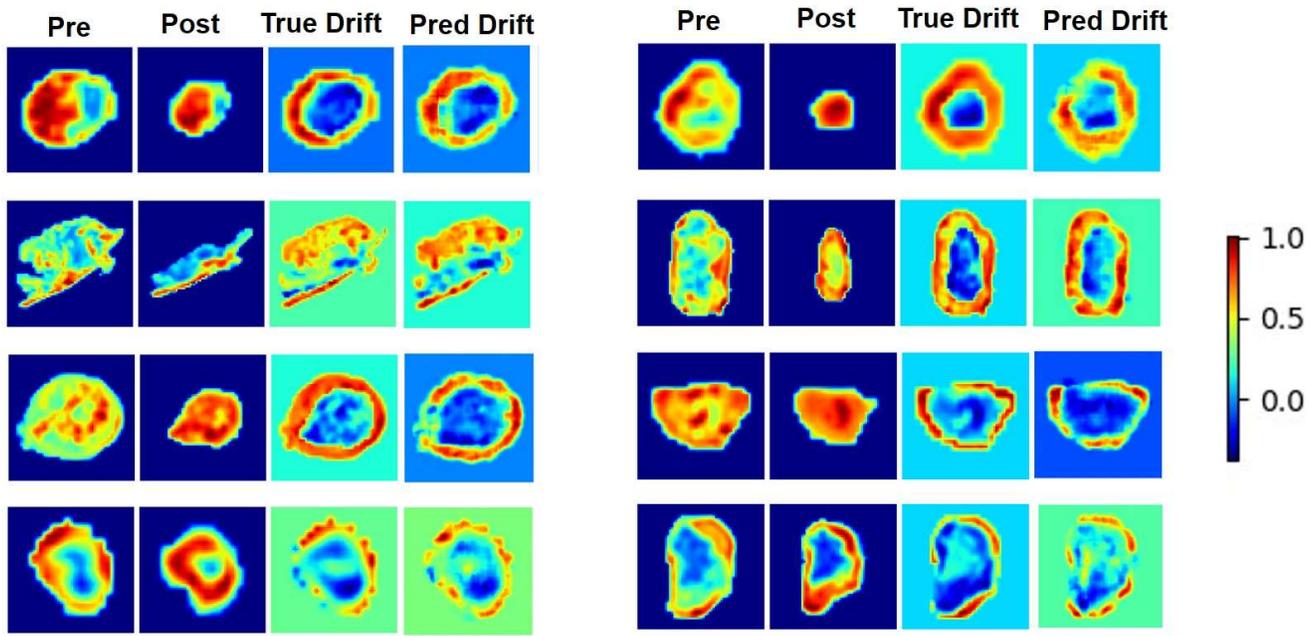

**Figure 8.** Performance of Method 2 for lesions showing good response (tumor volume reduction larger than 20%).